# Generation of cylindrically symmetric modes and orbital-angular-momentum modes with tilted optical gratings inscribed in high-numerical-aperture fiber

Liang Fang, Hongzhi Jia*, Hai Zhou, and Baiying Liu

*Engineering Research Center of Optical Instrument and System, Ministry of Education, Shanghai Key Lab of Modern Optical System School of Optical-electrical and Computer Engineering, University of Shanghai for Science and Technology, No. 516 JunGong Road, Shanghai 200093, China*

*Corresponding author: hzjia@usst.edu.cn

Optical fiber with high numerical aperture (NA) can efficiently relieve the degeneracy of higher-order linearly polarized (LP) modes. The degeneracy relief is investigated in two types of high NA fibers, i.e., low-index cladding fiber and high-index-core fiber. A naked-core fiber, as with low-index cladding, can be used theoretically to generate the orbital-angular-momentum mode (OAMM) $HE_{21}$ and the cylindrically symmetric modes (CSMs) $TM_{01}$ and $TE_{01}$. A high-index-core fiber incorporated with high-contrast-index structure can be used similarly to obtain OAMM $HE_{31}$. Both the generation of CSMs and OAMMs required tilted optical gratings to couple the fundamental core mode $HE_{11}$ into these modes. Tilt angle and modulation period of the grating fringes can be calculated simply and visually with the method proposed in this article. © 2014 Optical Society of America

## 1. INTRODUCTION

Optical modes with orbital angular momentum (OAM) characterized by the term exp $(il\phi)$ where $l$ is the topological charge number have attracted much attention, with applications including optical communications and manipulation [1-4]. Orbital-angular-momentum modes (OAMMs) multiplexing for date transmission in fiber has developed into a promising technology with the potential to massively increase the capacity of data transmission [5, 6]. As a special case, when $l = 0$ in above expression, cylindrically symmetric modes (CSMs) possess unique characteristics of polarization, including radially polarized $TM_{01}$ modes and azimuthally polarized $TE_{01}$ modes, and thus have been widely applied to electron acceleration [7], optical trapping [8], laser machining [9], tight focusing [10], three-Dimensional focus engineering, etc. [11, 12]. In a typical fiber, the OAMMs as vector modes are degenerated into linearly polarized (LP) modes, due to slight difference in effective refractive indices of at least two vector modes [13, 14]. Therefore, if these OAMMs can be produced in conventional fiber, it is beneficial to separate vector modes from the degenerated LP modes by enlarging the effective index differences between relevant vector modes. In precious reports, high-contrast-index ring-core-structure fiber was used to relieve mode degeneracy and generate OAMMs and CSMs [15-17]. In this article, we investigate another method of relieving mode degeneracy with high numerical-aperture (NA) optical fiber, and thus generating CSMs and OAMMs through tilted optical gratings.

A high-NA fiber with a large normalized waveguide frequency $V$ supports several higher-order core modes as vector modes, instead of linearly polarized (LP) modes. The conventional scalar mode $LP_{11}$ is split into OAMM $HE_{21}$ and either the CSMs $TM_{01}$ or $TE_{01}$, $LP_{21}$ is split into OAMMs $HE_{31}$ and $EH_{11}$, in other words, the differences in effective refractive indices between these relevant modes become distinct [14]. Optical fiber gratings with tilted modulation fringes can separately couple the fundamental core mode $HE_{11}$ into desired OAMMs, as well as CSMs $TM_{01}$ and $TE_{01}$. The tilted gratings formed in high-NA fiber can be implemented using the point-by-point and phase mask techniques with interferential ultraviolet exposure or femtosecond laser pulses [18, 19]. Both the period and tilt angle of grating fringes can be determined through a visualized calculation based on the phase-matching condition and the ratio between longitudinal and transverse modulation period, respectively. This calculation provides a new sight to understand the effect of grating tilt on the coupling coefficient, apart from the previous reports through the abstract equation [20, 21].

In this article, naked-core fiber, (i.e., air-core two layered cylindrical waveguides with NA over 1 and normalized waveguide frequency $V$ exceeding 18), is theoretically used to demonstrate a conversion from the fundamental core mode to the OAMM $HE_{21}$ and CSMs by using tilted optical gratings. In addition, a similar approach shows that a high-index-core fiber with a high-contrast-index structure is capable of generating OAMM $HE_{31}$. Naked-core fiber can be possibly fabricated by etching the cladding of a conventional fiber; high-index-core fibers may be similarly manufactured with As-Se/Ge-As-Se glasses as core/cladding waveguide materials [22], or with core/cladding materials reported in the Ref. [23]. Disregarding the weakened mechanical strength of naked-core fiber and the manufacturing challenges of these high-NA fibers, we focus on the analysis of relieving the mode degeneracy with the two types of high-NA fibers, such as a naked-core fiber and a high-index core fiber, and then theoretically generate CSMs and OAMMs $HE_{21}$ and $HE_{31}$ with mode coupling through tilted optical gratings in this article. This kind of all-fiber generation method may result in good compatibility, high conversion efficiency and purity, and simple construction, compared to the methods of generating and multiplexing OAMMs with spatial-light modulators or other free-space components and beam splitters [24, 25, 4], which may allow for practical generation and conversion of CSMs and OAMMs.

## 2. ANALYSIS OF RELIEVING THE MODE DEGENERACY

The division of mode degeneracy and modal dispersion are discussed in this section. The two types of high-NA fiber mentioned earlier are used for our theoretical analysis (naked-core fiber where the cladding around the general fiber is stripped, and high-contrast-index structure known as high-index-core fiber). Dispersion equations of the core modes of two-layer waveguides in fiber are listed as follows [13]:

for hybrid mode $HE_{vm}/EH_{vm}$,

$$\left[\frac{J'_v(u_1 a_1)}{u_1 J_v(u_1 a_1)} + \frac{K'_v(w_2 a_1)}{w_2 K_v(w_2 a_1)}\right]\left[\frac{J'_v(u_1 a_1)}{u_1 J_v(u_1 a_1)} + \frac{n_2^2}{n_1^2}\frac{K'_v(w_2 a_1)}{w_2 K_v(w_2 a_1)}\right]$$
$$= -\left(\frac{jv\beta}{k_0 n_1 a_1}\right)^2 \left(\frac{1}{u_1^2} + \frac{1}{w_2^2}\right)^2 \quad (1)$$

for mode $TE_{01}$,

$$\frac{K'_0(w_2 a_1)}{w_2 K_0(w_2 a_1)} + \frac{J'_0(u_1 a_1)}{u_1 J_0(u_1 a_1)} = 0, \quad (2)$$

and for mode $TM_{01}$,

$$\frac{K'_0(w_2 a_1)}{w_2 K_0(w_2 a_1)} + \frac{n_1^2}{n_2^2}\frac{J'_0(u_1 a_1)}{u_1 J_0(u_1 a_1)} = 0, \quad (3)$$

where $u_1 = k_0\sqrt{n_1^2 - n_{eff}^2}$ and $w_2 = k_0\sqrt{n_{eff}^2 - n_2^2}$, $k_0 = 2\pi/\lambda$ and $\beta = k_0 n_{eff}$ present the wave vector and the propagation constant of light, respectively; $n_1, n_2$ and $n_{eff}$ are the refractive indices of the inner and outer layer waveguides, and the effective indices of guided mode, respectively; $a_1$ is the core radius, and the integers $v$ and $m$ denote the mode ordinal and corresponding radial ordinal.

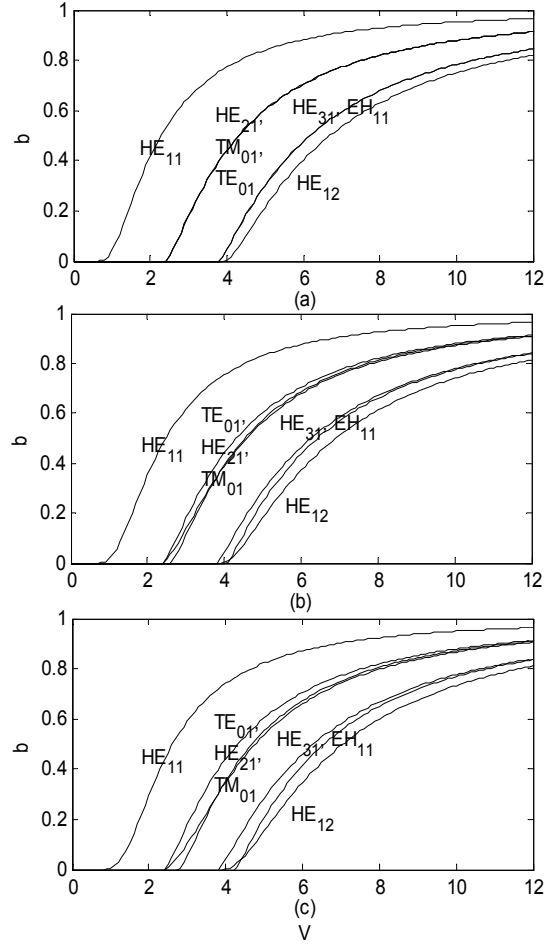

Fig. 1 Modal dispersion curves of several vector modes, (a) ordinary SMF, $n_1 = 1.4681$, $n_2 = 1.4628$; (b) high-index-core fiber $n_1 = 1.85$, $n_2 = 1.46$; (c) naked-core fiber, $n_1 = 1.4681, n_2 = 1.00$.

The fiber parameters of the high NA-fibers are $n_1 = 1.4681$ and $n_2 = 1.00$ for a naked-core fiber, $n_1 = 1.85$ and $n_2 = 1.46$ for a high-index-core fiber, both with a core radius $a_1 = 4.15$ µm. The dispersion curves of several vector modes including $HE_{11}$, $HE_{21}$, $TM_{01}$, $TE_{01}$ $HE_{31}$, $EH_{11}$, and $HE_{12}$ are depicted in Fig. 1(a) for an ordinary single-mode fiber (SMF), those of a high-index-core fiber are depicted in Fig. 1(b), and those of a naked-core fiber are depicted in Fig. 1(c). In the figure, the variable $b = (n_{eff}^2 - n_2^2)/(n_1^2 - n_2^2)$ indicates the normalized effective refractive index, $V = k_0 a_1\sqrt{n_1^2 - n_2^2}$ is the normalized waveguide frequency that becomes a function depending solely on the operating wavelength $\lambda$ and the core radius $a_1$. Note that the dispersion curves of other higher-order vector modes, if supported, are not included in Fig. 1. In general, it can be seen that the dispersion curves of OAMMs $HE_{21}$, $HE_{31}$ and $EH_{11}$, CSMs $TM_{01}$ and $TE_{01}$ are almost completely split in high NA fibers shown in Fig. 1(b) and 1(c), compared with those in Fig. 1(a). This situation can be explained by the large ratio between $n_1^2$ and $n_2^2$ in Eqs. (1) and (3); in other words, the splitting is caused by a high $NA = \sqrt{n_1^2 - n_2^2}$, which reaches 1.14 for a high-index-core fiber and 1.07 for a naked-core fiber, but only 0.12 for the SMF.

In order to reveal the effective index differences between the fundamental core mode $HE_{11}$ and each vector mode under various $NA$ for both high-index-core fiber and low-index-cladding fiber, which directly determines the dispersion degree for mode coupling from $HE_{11}$ to other modes, according to Eqs. (1), (2) and (3), we define a function of the effective index $n_{eff}$ relative to $NA$,

$$n_{eff}^i = f_i(NA), \quad (4)$$

with $i=11$, for $HE_{11}$; $i=21$, for $HE_{21}$; $i=m01$, for $TM_{01}$; for $TE_{01}$, $i=e01$, for $HE_{31}$, $i=31$; for $EH_{11}$, $i=h11$; for $HE_{12}$, $i=12$. Then the $\Delta n_{eff}$ of mode couples stemming from $HE_{11}$, combined with $HE_{21}$, $TM_{01}$, $TE_{01}$, $HE_{31}$, $EH_{11}$, and $HE_{12}$, can be expressed as follows:

$$\Delta n_{eff}^{11-i} = f_{11} - f_i. \quad (i \neq 11) \quad (5)$$

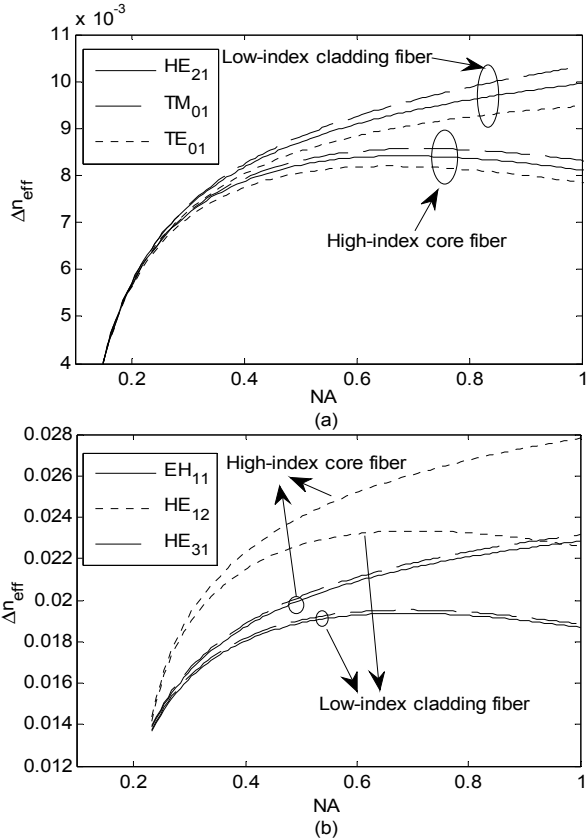

Fig. 2 Relationships between different $\Delta n_{eff}$ under various $NA$ for $HE_{21}$, $TM_{01}$, $TE_{01}$ (a), and $HE_{31}$, $EH_{11}$, and $HE_{12}$ (b) in both high-index core and low-index cladding fiber.

In Fig. 2(a), three kinds of $\Delta n_{eff}^{11-i}$ for $HE_{21}$, $TM_{01}$, $TE_{01}$ are illustrated for high-index-core fiber with a fixed index $n_2 = 1.46$; and for low-index-cladding fiber with a fixed index $n_1 = 1.46$, both under the parameters of $a_1 = 4150$ nm and $\lambda = 1550$ nm. It can be seen that the three $\Delta n_{eff}$ are obviously separated with increasing $NA$, and the degree of separation in the low-index-cladding fiber is more distinct than in the high-index-core fiber. Similarly, the other three kinds of $\Delta n_{eff}^{11-i}$ for $HE_{31}$, $EH_{11}$, and $HE_{12}$ are shown in Fig. 2(b). We can see that the effect of $NA$ on separation of $\Delta n_{eff}^{11-31}$ and $\Delta n_{eff}^{11-h11}$ is slight when $NA$ increases beyond a certain point.

## 3. MODE COUPLING WITH TILTED OPTICAL GRATINGS

In this section, naked-core fiber and high-index-core fiber are used to achieve mode conversion from the fundamental core mode $HE_{21}$ to CSMs $TM_{01}$ and $TE_{01}$, and OAMMs $HE_{21}$ and $HE_{31}$ using tilted optical gratings. Based on the coupled-mode theory [26], coupling via an optical fiber grating must satisfy two requirements, the phase-matching condition must be met, and a large coupling coefficient between the two related modes is required. For the phase-matching condition of transmissive optical gratings,

$$\frac{2\pi}{\lambda}\Delta n_{eff} = \frac{2\pi}{\Lambda_0}, \quad (6)$$

where $\Lambda_0$ is the grating period, and $\Delta n_{eff}$ is the difference in the two modes' effective refractive indices. For obtaining CSMs and OAMM $HE_{21}$ with naked-core fiber, we substitute the independent variable $NA$ for $V/NA$ where $NA = 1.07$, while for OAMM $HE_{31}$ with high-index core fiber, $NA = 1.14$, the function in Eq. (4) then becomes

$$n_{eff}^i = f_i(V/NA), \quad (7)$$

Inserting Eqs. (5) and (7) into Eq. (6), and multiplying the new equation by core radius $r$ on both sides, yield a new expression involving the variable $\Lambda, r$ and $\lambda$

$$\frac{2\pi r}{\Lambda_0} = \frac{2\pi r}{\lambda}\left[f_{11}\left(\frac{2\pi r}{\lambda}\right) - f_i\left(\frac{2\pi r}{\lambda}\right)\right]. \quad (i \neq 11) \quad (8)$$

The relevant expression in Eq. (8) plays a significant role on determining the grating period in the case of different core radius and operating wavelength.

The coupling coefficient is expressed by [27]

$$\kappa_{11-i} = \frac{1}{2}\omega\varepsilon_0 n_1^2 \bar{\delta} \int_0^{2\pi} d\phi \int_0^{a_1} \exp(Kr\cos\phi\tan\theta) E_{11} E_i^* r dr, \quad (9)$$

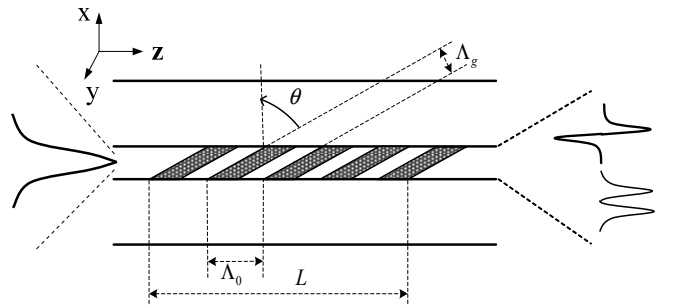

Fig. 3 Structure diagram of tilted optical gratings in a high-NA fiber, and the coupling process from the fundamental mode $HE_{11}$ (left) to vector modes $TM_{01}$, $TE_{01}$, $HE_{21}$ or $HE_{31}$ (right).

where $\omega$ and $\varepsilon_0$ are the angular frequency and the dielectric constant in vacuum, $\bar{\delta}$ is the average modulation strength, $E_{11}$ and $E_i$ are functions of electric field transverse distribution of the mode $HE_{11}$; $HE_{21}$ when $i = 21$; $TM_{01}$ when $i = m01$; $TE_{01}$ when $i = e01$, $HE_{31}$ when $i = 31$; $EH_{11}$ when $i = h11$; and $HE_{12}$ when $i = 12$, the superscript $*$ indicates a complex conjugate, $K = 2\pi/\Lambda_0$ is the grating constant, and $\theta$ is the tilt angle of gratings.

The structure diagram of tilted optical gratings written in high-$NA$ fibers including a naked-core finer and a high-index-core fiber are demonstrated in Fig. 3, where a mode conversion is observed from the fundamental mode $HE_{11}$ to CSMs $TM_{01}$ and $TE_{01}$, and OAMMs $HE_{21}$ and $HE_{31}$. Note that the field distributions are described by the electric field intensities with direction indication. In Eq. (9), with tilted gratings, it is the transverse modulation period $\Lambda_x = \Lambda_0/\tan\theta$ that acts on the expression for the coupling coefficient where $r\cos\phi$ implies the displacement in the direction of transverse modulation.

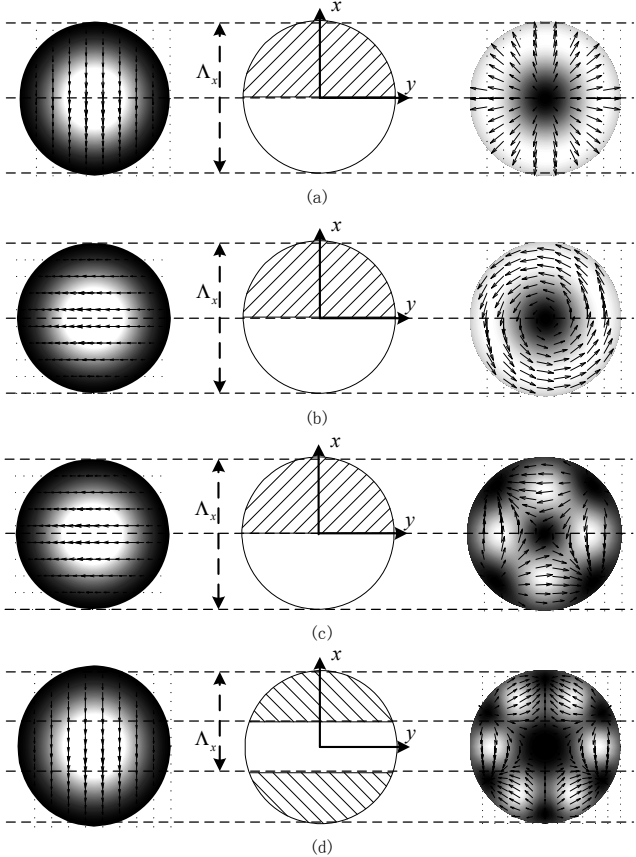

Fig. 4 Relationships between transverse modulation period $\Lambda_x$ and mode coupling from $HE_{11}$ to $TM_{01}$ (a), to $TE_{01}$ (b), to $HE_{21}$ (c), and to $HE_{31}$ (d).

To maximize the integral value when efficiently obtaining CSMs and OAMM $HE_{21}$ with a naked-core fiber, the maximum coupling coefficient from $HE_{11}$ to these modes occurs with a transverse modulation period $\Lambda_x \approx 2a_1$. The relationship between the field distributions of $HE_{11}$ coupled with CSMs and OAMM $HE_{21}$ are illustrated in Fig. 4(a)-(c), along with the transverse modulation fringes. In conventional fiber, i.e., a weakly guiding fiber, $HE_{21}$ coupled from $HE_{11}$ will be degenerated into $LP_{11}$ with $TM_{01}$, or into its cross-polarization $LP_{11}$ with $TE_{01}$, which has been previously reported [28]. In our previous work, we have revealed the mode coupling between two core modes $LP_{0m}$, as well as between higher-order core modes $LP_{0m}$ and cladding modes $HE_{1m}$ [29, 30]. Here, for the tilted gratings, the tilt angle $\theta$ and actual grating period $\Lambda_g$ can be visually calculated by

$$\theta = \tan^{-1}\left(\frac{\Lambda_0}{\Lambda_x}\right). \qquad (10)$$
$$\Lambda_g = \Lambda_0 \cos\theta$$

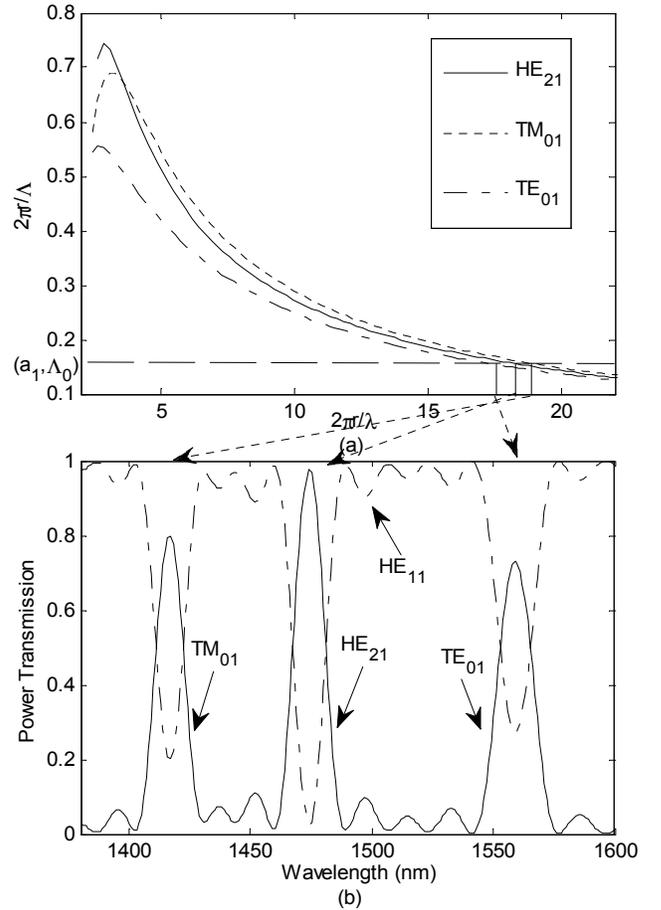

Fig. 5 (a) Curves of $2\pi r/\Lambda$ versus $2\pi r/\lambda$ for CSMs and OAMM $HE_{21}$ coupled from $HE_{11}$ in a naked-core fiber. (b) Power exchange spectrum from $HE_{11}$ to these three modes.

According to Eqs. (1)-(3), (7) and (8), the curves of $2\pi r/\Lambda$ versus $2\pi r/\lambda$ expressed by Eq. (8) for CSMs and OAMM $HE_{21}$ are depicted in Fig. 5(a). From this, the grating period $\Lambda_0$ can be determined for desired CSMs and OAMM $HE_{21}$ at any operating wavelength. Then the actual grating period $\Lambda_g$ and tilt angle can be obtained based on Eq. (10). Here if power exchanges of these three modes fall within the wavelength range of 1400-1580nm, and the core radius of fiber are taken as 4150nm, the corresponding actual grating period $\Lambda_g$ and tilt angle can be calculated (see Table 1). This table also contains the normalization coupling coefficients based on Eq. (9) and the product between the optional grating length and modulation strength in the case of maximum coupling efficiency determined by $\eta = \sin^2(\kappa L/2) = 1$ [26]. Power conversion spectrum from $HE_{11}$ to CSMs and OAMM $HE_{21}$ are depicted in Fig. 5(b). Note that the input polarization of the fundamental core mode $HE_{11}$ for the wavelength of 1420 nm needs to be orthogonal to the wavelength of 1560 nm, which is expounded in Fig. 4(a) and 4(b). From the Fig. 5(b), we can see that the vector modes $HE_{21}$, $TM_{01}$ and $TE_{01}$ are completely split in the transmission spectrum, and have high efficiency of power exchanges. Actually, the efficiency can be controlled by adjusting modulation strength or the grating length for individual CSM $TM_{01}$ or $TE_{01}$ or OAMM $HE_{21}$.

*Tab. 1 Grating parameters of tilted optical gratings and corresponding coupling coefficients*

| $\lambda\ (nm)$ | 1417 | 1475 | 1560 | 1537 |
|---|---|---|---|---|
|  | $TM_{01}$ | $HE_{21}$ | $TE_{01}$ | $HE_{31}$ |
| $\Lambda_g\ (um)$ | 8.29 | | | 5.52 |
| $\theta$ | 87.08° | | | 86.30° |
| $\kappa/\bar{\delta}\ (um^{-1})$ | 2.09 | 2.69 | 1.95 | 2.17 |
| $\bar{\delta}L\ (um)$ | 1.05 | | | 1.45 |

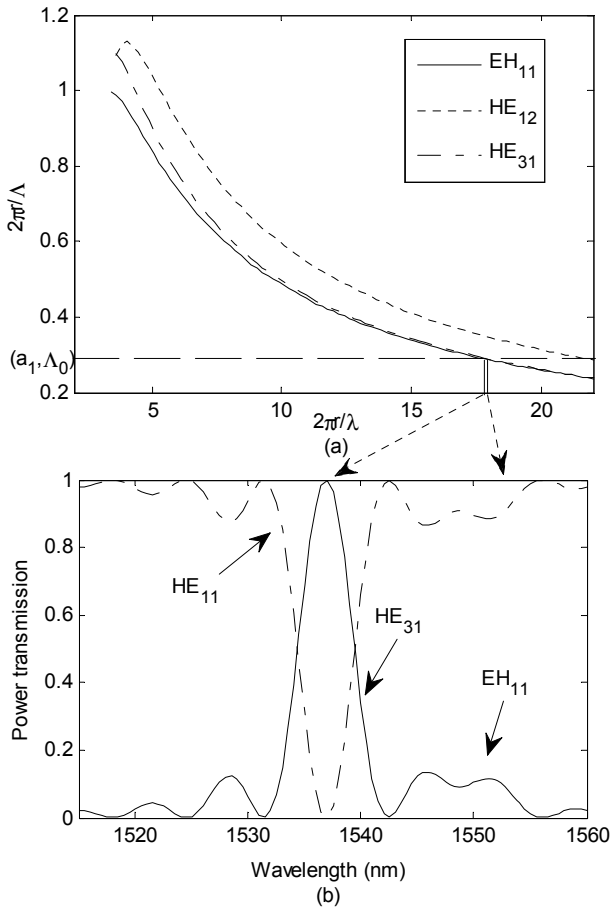

Fig. 6(a) Curves of $2\pi r/\Lambda$ versus $2\pi r/\lambda$ for OAMM $HE_{31}$, $EH_{11}$, and $HE_{12}$ coupled from $HE_{11}$ in a high-index core fiber. (b) Power exchange spectrum from $HE_{11}$ to OAMM $HE_{31}$.

Similarly, to obtain OAMM $HE_{31}$ for a high-index-core fiber, while maximizing the coupling coefficient, the transverse modulation period should be $\Lambda_x \approx 4a_1/3$. Fig. 4(d) illustrates the relationship between the field distributions for $HE_{11}$, OAMM $HE_{31}$ and the transverse modulation fringes. The curves of $2\pi r/\Lambda$ versus $2\pi r/\lambda$ expressed by Eq. (8) for OAMMs $HE_{31}$, $EH_{11}$, and $HE_{12}$ with second radial order are depicted in Fig. 6(a). The parameters calculated similarly are listed in Table 1. Power conversion spectrum from $HE_{11}$ to OAMM $HE_{31}$ is presented in Fig. 6(b). Note that for this spectrum, mode $EH_{11}$ has a low conversion rate because of the small coupling coefficient owing to the transverse modulation period shown in Fig. 5(d). Moreover, the point of power conversion from $HE_{11}$ to $HE_{21}$ is beyond the wavelength range of the spectrum in Fig. 6(b). Other higher-order OAMMs can be converted from the fundamental core mode $HE_{11}$ by similar methods, provided the gratings' longitudinal modulation period meets the phase-matching condition. Meanwhile, the transverse modulation fringes are adjusted to make the coupling coefficients between $HE_{11}$ and desired OAMMs sufficiently large.

## 4. SUMMARY

High-NA fiber has been studied in this article as a means for generating and transmitting CSMs and OAMMs. The NA can be greatly increased by increasing or decreasing a fiber's core and cladding indices. Mode degeneracy is relieved in the high-NA fiber so that the scalar mode $LP_{11}$ is divided into CSMs and OAMM $HE_{21}$, and $LP_{21}$ mode is divided into OAMMs $HE_{31}$ and $EH_{11}$. These CSMs and OAMMs can be generated by tilted optical gratings inscribed in this high-NA fiber. In this article, naked-core fiber was used with such gratings to achieve theoretical generation of CSMs and OAMM $HE_{21}$, coupled with the fundamental core mode $HE_{11}$. Similarly, OAMM $HE_{31}$ is obtained with tilted optical gratings inscribed in high-index-core fiber, and this method may be expanded for the generation of other higher-order OAMMs.

## ACKNOWLEDGEMENT

This work is supported by the National Basic Research Program of China (2013CB707500), the Innovation Fund Project for Graduate Student of Shanghai (JWCXSL1302), and partly supported by the Shanghai Leading Academic Discipline Project (No.S30502).

## REFERENCS